\documentclass[aps,prl,epsfig,showpacs,floats,twocolumn,amssymb,amsmath,floatfix,
groupedaddress,superscriptaddress]{revtex4-1}

\usepackage{graphicx} 
\usepackage{dcolumn} 
\usepackage{bm} 
\usepackage{float} 
\usepackage{bbm} 
\usepackage{color}
\usepackage{xcolor}
\usepackage{amsfonts}
\usepackage{lmodern} 
\usepackage{amsmath} 
\usepackage{amsmath,amssymb}
\begin{document} 
\title{Instabilities in a growing system of active  particles: scalar and vector systems}

\author{Forouh Maleki} 
\affiliation{Department of Physics, Institute for Advanced Studies in Basic Sciences 
(IASBS), Zanjan 45137-66731, Iran} 

\author{Ali Najafi} 
\email[]{najafi@iasbs.ac.ir}
\affiliation{Department of Physics, Institute for Advanced Studies in Basic Sciences 
(IASBS), Zanjan 45137-66731, Iran} 
\affiliation{Research Center for Basic Sciences \& Modern Technologies (RBST), Institute for Advanced Studies in Basic Sciences, Zanjan, Iran}

\date{\today} 
\begin{abstract} 
The physics of micron-scale biological colonies usually benefits from different out-of-equilibrium sources. In bacterial colonies and cellular tissues, the growth process 
is among the important active sources that determine the dynamics. In this article, we study the generic dynamical instabilities associated with the growth phenomena that may arise in both scalar and vectorial systems. In vectorial systems, where the rotational degrees of particles play a role,  a phenomenological growth-mediated torque can affect the rotational dynamics of individual particles. We show that such a growth-mediated torque can result in active traveling waves in the bulk of a growing system. In addition to the bulk properties, we analyze the instabilities in the shape of growing interfaces in both scalar and vectorial systems.
\end{abstract} 
\maketitle

\section{Introduction.}
The process  of growth is a necessary element that brings the meaning of life to the living systems.  
From a physicist's standing point, 
one important and central challenge lies in understanding the mechanism by which a non-equilibrium proliferating system forms its overall functioning shape \cite{cowin2004tissue,newman1990generic}.  
Bacterial colonies \cite{orozco2013order,cicuta,PhysRevLett.117.048102}, biofilms \cite{van2002mathematical,raminbiofilm,beroz2018verticalization}, and growing tissues \cite{growtissue1,growtissue2,growtissue3,basan2011undulation} are 
standard examples that belong to the class of active systems where 
one can study the growth phenomena. Along this general task, self-organization and ordering in active colonies \cite{vicsek1995novel,tsimring1,mahadevan1,philipbittihn,giomi1,alert2020physical}, pattern formation in biological systems 
\cite{gierer1972theory,family1987deterministic,vicsek1995novel} and nematic ordering in bacterial colonies \cite{drescher2016architectural,zhang2010collective} are studied extensively.

A growing system benefits from chemical, physical, and biological 
processes at many different time and length scales \cite{stooke2018physical}. 
On the other hand, different mechanisms ranging from behavior at the level of individual cells,  
cell-cell signaling, and environmental feedback, help a growing system to perform its job. 
All of such processes are 
mostly based on non-equilibrium reactions that eventually aim to provide mechanical motion. 
In a simplified mesoscale mechanical picture, out-of-equilibrium forces can be modeled by active terms in a phenomenological description that is called the active nematics \cite{bacterialcolonynature,doostmohammadi2,doostmohammadi1}. Such continuum descriptions accompanied by agent-based simulations have to be compared with experimental facts. 
 Resulting from nonlinearities hidden in continuum  
models, physical instabilities are 
among the intriguing phenomena that can help the system to find its overall shape.  
Examples of such instabilities include the buckling at bulk and roughening at boundaries  \cite{tsimring2,karsten1,basan2011undulation,julicher1,ricard1,datta,raminbiofilm}.
 
 In this article, we aim to present a generic description of a growing active matter that takes into account the growth 
  at a phenomenological level. To put our idea, we will consider a growing matter in two categories 
 of scalar and vectorial cases. In a scalar system, the rotational degrees of freedom of individual cells are neglected while in the vector case, the rotational motion plays an important role.  In the vectorial case, in addition to the density, the director field is also a relevant variable that needs to take into account.  Based on symmetry arguments, we consider a  growth-mediated torque in our description
and investigate the instabilities in both bulk and boundaries of a growing system. To this end, we use a minimal model that can capture the mechanics of a growing system.

\section{Model}
\begin{figure}[h!] 
 \begin{center}
 \includegraphics[width=0.45\textwidth]{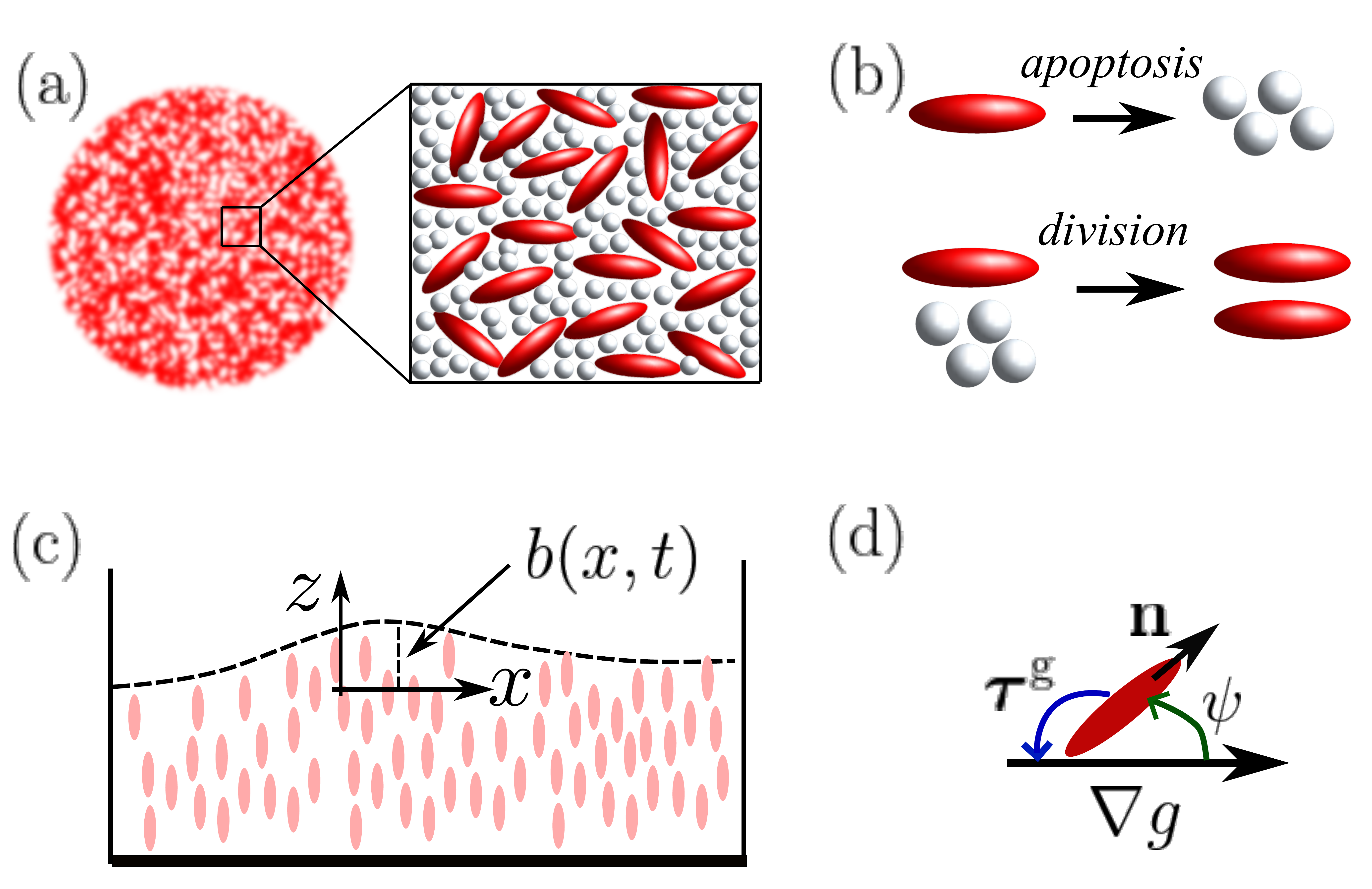}
      \caption{(a) Schematic view of a proliferating system composed of anisotropic active particles moving in an ambient fluid.
      (b) Detailed processes of apoptosis and cell division. (c) A growing front with fluctuating shape. (d)  Growth-mediated torque tends to align a particle with the direction of the growth gradient.}
\label{fig1} 
  \end{center}
\end{figure}

As shown in Fig.~\ref{fig1} (a,b), consider a two-dimensional system composed of motile particles with the proliferation ability. 
This dense system of active particles lives in an ambient fluid, a fluid that could be either an 
aqueous  media (in bacterial suspension) or extracellular fluid (in 
growing tissue).
For biological systems, apoptosis and cell division can contribute and result in a positive or negative overall growth rate. 
We denote by $g(t)$, the rate by which the particles proliferate.  In addition to growth, the motility of particles would also 
act as another source for initiating
mechanical motion in our system. Each self-driven motile particle can exert stress on the fluid. 
Denoting by $a$, the amount of stress that each active particle carries,   this stress could be either positive or negative. 
Extensile (pusher) and contractile (puller) active particles will be described by  $a > 0$ and $a<0$, respectively.
Two phenomena of active 
motility of particles and the process of proliferation can result in large-scale motion in this system. In general, the time scale for the 
motion in the active part of the system may be much smaller than its counterpart in the ambient fluid. 
As a result of this observation and to study the long-time behavior of the system, we only consider the dynamics of the 
active part. At the continuum level   
the physical state of this system can be described by the coarse-grained fields of density $\rho$,  velocity filed ${\bf v}({\bf r},t)$ and director field $\mathbf{n}(\mathbf{r},t)$ of the active part. These fields are subjected to the following dynamical equations \cite{prost,chaikin1995principles}:
\begin{eqnarray}
&&\rho\frac{d}{dt}{\bf v}=\nabla\cdot\Sigma-\Gamma {\bf v},~~~~\nabla\cdot{\bf v}=g(t),\nonumber\\
&&\frac{D}{Dt}{\mathbf{n}}=\gamma^{-1}\left({\bf h}+{\bf n}\times\boldsymbol{\tau}^\text{g}\right),
\label{eqm}
\end{eqnarray}
where co-moving and co-rotating derivatives are defined as: $\frac{d}{dt}=\partial_t+{\bf v}\cdot \nabla$ and 
$\frac{D}{Dt}=\frac{d}{dt}+(I-{\bf n}{\bf n})\cdot {\cal D}\cdot {\bf n} $, respectively. 
Here $I$ denotes the unit tensor of rank two and 
${\cal D}=D^-+AD^+$ with $D^\pm=(1/2)(\nabla{\bf v}\pm[\nabla{\bf v}]^\text{T})$. For spherical particles $A=0$ and oblate (prolate) particles correspond to $A>0$ ($A<0$) \cite{jeffery}.
In our active system, both growth-mediated torque  $\boldsymbol{\tau}^\text{g}$, a phenomenological term that we will introduce later and 
the thermodynamic current $\Sigma$  derive the system to out of equilibrium conditions. 
The thermodynamic force can be written as \cite{chaikin1995principles}: 
\begin{equation}
\Sigma= \Sigma^\text{d} + \Sigma^\text{r} -\frac{\partial{\cal F}}{\partial \nabla \mathbf{n}}
\cdot \nabla \mathbf{n}-a{\bf n}{\bf n}-PI.
\end{equation}
The term proportional to 
the local nematic tensor ${\bf n}{\bf n}$ is the active stress resulting from the motility of particles \cite{ramaswamy1}. 
Furthermore, the elastic free energy density of the  nematic phase and corresponding molecular force can be written as \cite{prost, landau}:
\begin{equation}
{\cal F}= 
\frac{K}{2}\sum_{i,j} \partial_i n_j\partial_i n_j, ~~~~{ h}_i=-{\delta{\cal F}}/{\delta  {n}_i}.\nonumber
\end{equation}
Recalling the equation of motion (Eq.~\ref{eqm}), the rotational friction coefficient is denoted by $\gamma$ and the term proportional to $\gamma$ guarantees a relaxation to states with zero molecular force in passive systems.  
The dissipative part of stress tensor is given by $\Sigma^\text{d}=2{\eta} [D^+-1/2(\nabla\cdot{\bf v}){ I}]$  and 
reactive stress is given by: $\Sigma^\text{r}=\frac{1}{2}(\mathbf{n}\mathbf{h}- \mathbf{h} \mathbf{n})- \frac{A}{2}(\mathbf{n}\mathbf{h}+\mathbf{h}\mathbf{n})$. Friction with the substrate is denoted by a single parameter
$\Gamma$. 

It is important to note that the number density of such a growing system is not conserved. In this 
case, a detailed model for the pressure should be considered.    
Growth pressure denoted by $P$ is the main place where the growth affects the dynamics.  
Following the well-studied two-fluid model \cite{julicher2,julicher3}, we choose a simple model for this growth pressure. 
For active systems similar to biological tissues, the concept of homeostasis works. 
In such systems, the static steady-state can be described by a characteristic homeostatic pressure denoted by $P_\text{h}$.    
Slowly growing systems are very near to the homeostatic state and the pressure can be expanded as powers of growth rate. It can be shown that in this regime, density $\rho$ can be considered approximately as a constant variable. Furthermore, the pressure obeys the relation \cite{julicher2}:
\begin{equation}
P=P_\text{h}-\zeta\nabla\cdot{\bf v},
 \end{equation}
where the bulk viscosity for the homeostatic state is denoted by $\zeta$. For $\zeta>0$, any local increase in the pressure would result in the domination of apoptosis over cell division.  

It should be noted that the low Reynolds condition that is relevant for our purposes, will exclude any non-linearity in terms of  
velocity field emerging from co-moving derivatives. Nutrients are also assumed to be accessible everywhere without any limitations. 
This last simplification can work well for 2-D colonies where the third dimension always provides free space to supply the food. 

How does the growth affect the orientational degree of freedom? 
 As a result of short-range cell-cell communications, 
the growing state 
 of the cells that are surrounding a target cell, can influence the motion of this target cell. This will result in a 
 growth-mediated torque that will be denoted by $\boldsymbol{\tau}^\text{g}$.
 The existence of such torque was discussed previously \cite{bacterialcolonynature}. 
 The polarity and geometrical asymmetry of the particles 
 might influence this scenario. In a phenomenological description, the local gradient of growth rate can contribute to the torque 
 $\boldsymbol{\tau}^\text{g}$, that is exerted on the cells. 
At the leading order of the growth gradient, and following the symmetry considerations, 
a term like  $\boldsymbol{\tau}^\text{g}\sim \nabla g \times{\bf n}$ is a possible term that we will consider.  
In the homeostatic picture, the growth rate is proportional to the pressure, then the growth-mediated torque will be written as:
\begin{equation}
\boldsymbol{\tau}^\text{g}=
-\beta A^2 \nabla P \times{\bf n},
\end{equation}
where, $\beta$ is a phenomenological parameter, and a simple second-order dependence on the asymmetry parameter $A$ is assumed, meaning that the torque is similar for oblate and prolate particles. 
For $\beta>0$, as it is shown in fig.~\ref{fig1}(d), the growth torque tends to align the particles' polarity and $-\nabla g$. 
Recalling the angle between ${\bf n}$ and $\nabla g$ by $\psi$, for a fixed growth gradient, $\psi =\pi$ is an absorbing state, meaning that the particle tends to move toward a region with less growing rate.

The model described so far has many ingredients describing different physical processes that can affect the dynamics. 
We try to consider the 
effects of different terms step by step. First, we consider a scalar model in which we neglect the orientational degrees of freedom of the particles by dropping out the variable 
${\bf n}$. Later on, we will add the effects of the nematic variable ${\bf n}$.

\section{Scalar system}
To study the bulk properties of a growing scalar system, we first consider an unbounded growing system. 
For a gradually growing system in which the growth rate is very small, we linearize the equations in terms of the velocity field ${\bf u}({\bf r},t)$ and study its dynamics. 
Defining the Fourier transform 
of any variable $f_i({\bf r},t)$ as ${\tilde f}_i({\bf q},\omega_i)=\int d{\bf r}dt f_i({\bf r},t)\exp{[i({\bf q}\cdot{\bf r}-\omega_i t)]}$, 
we can observe that the dispersion for longitudinal and transverse modes obeys the following relations:
\begin{equation}
i\omega_{\text{L,T}}=\rho^{-1}\left(\Gamma+(\eta+A_{\text{L,T}}\zeta)q^2\right),
\end{equation} 
where $A_{\text{L}}=1$
and $A_{\text{T}}=0$. Transverse and longitudinal directions are defined with respect to the direction of wave vector ${\bf q}$.  
 As it is seen, for this scalar case, any kind of fluctuation in the bulk will eventually disappear. This result roots in the homeostatic description where growth-mediated motion is assumed to propagate through the pressure variations. For this scalar case, all motions are expected to take place at the boundaries. For this reason, it is necessary to analyze the effects at the boundaries. 

To study the boundary effects, we consider the case where the system is allowed to grow in 
1-dimension. In this case, the system is limited from one side by a rigid wall and it is free 
to expand from the other side. As depicted in fig.~\ref{fig1}(c), we choose a reference frame 
with $z$-axis along the growth direction. The growing front lies at  $z=0$ while the limited part of the system sits at $z=-\infty$. 
The growing front is assumed to be a permeable and abrupt boundary (at $z=0$). 
On top of this permeable boundary $(z>0)$, a fluid reservoir with fixed pressure is in contact with the growing system. The upper fluid is in 
mechanical equilibrium with the ambient fluid at the tissue (part of the growing system that is passive with no dynamics in our model). 
Furthermore, we assume that an additional external mechanical pressure denoted by $P^\text{ext}$, is exerted on the active part of 
the system at the position of the boundary \cite{julicher2}. 
This externally applied pressure can help us to capture the physics of growth and homeostasis in living systems. 
The homeostatic state is a state in which the external pressure is adjusted to a specific value $P_\text{h}$ so that the boundary 
reaches a non-moving still state. 
In this case, the death and apoptosis processes cancel each other and the growth rate vanishes on average.
Obviously, any deviation from the homeostatic pressure will result in a motion in the boundary. Denoting by 
$P^\text{ext}= P_\text{h}+ \Delta P$, for 
$\Delta P<0$, the division dominates over death and the boundary will move upward. 

To consider the dynamics, we notice that the stress tensor for this scalar system reads as:
\begin{equation}
\Sigma=\eta(\nabla{\bf v}+\nabla{\bf v}^T)-\eta^-\nabla\cdot{\bf v}I-P_hI,
\nonumber
\end{equation}
and the dynamical equation $\nabla\cdot\Sigma=\Gamma{\bf v}$ takes the following form:
\begin{eqnarray}
&&(\eta\partial_{z}^{2}+\eta^+\partial_{x}^{2})v_x+\zeta\partial_x\partial_z v_z=\Gamma v_x,\nonumber\\
&&(\eta^+\partial_{z}^{2}+\eta\partial_{x}^{2})v_z+\zeta\partial_x\partial_z v_x=\Gamma v_z,\nonumber\\
\end{eqnarray}
where $\eta^\pm=(\eta\pm\zeta)$.
To solve these equations, we decompose the velocity field into two parts:
\begin{equation}
{\bf v}=(v_{z}^{s}(z)+\delta v_z)\hat z+\delta v_x{\hat x},
\end{equation}
where the steady state solution $v_{z}^{s}(z)$ corresponds to a steady state growth with a flat boundary.  The rest shows the possible fluctuations corresponding to time-dependent nonuniformity in the shape of the boundary.  Denoting the fluctuating shape of the 
boundary by function $b(x,t)$ (see fig.~\ref{fig1}), the velocity satisfies the following boundary condition:
\begin{equation}
v_z(x,b)-v_z(x,0)=\partial_tb(x,t)+v_x(x,b)\partial_xb(x,t).
\label{BC1}
\end{equation}
Denoting the surface tension of the growing boundary by $\gamma_s$, the  components of the stress tensor should satisfy the following 
relations:
\begin{eqnarray}
&&\Sigma_{nn}(x,z=b(x,t))=-P_h-\Delta P-\gamma_s\nabla\cdot{\hat n},\nonumber\\
&&\Sigma_{nt}(x,z=b(x,t))=0,
\label{BC2}
\end{eqnarray}
where ${\hat t}\approx {\hat x}$ and ${\hat n}\approx{\hat z}-\partial_xb(x,t){\hat x}$ are local tangent and normal vectors. 
In terms of Cartesian components, the stress tensor can be written as:
\begin{eqnarray}
&&\Sigma_{nn}=\Sigma_{zz}-2(\partial_xb)\Sigma_{zx},\nonumber\\
&&\Sigma_{nt}=\Sigma_{zx}+(\partial_xb)(\Sigma_{zz}-\Sigma_{xx}),
\end{eqnarray}
with,
\begin{eqnarray}
&&\Sigma_{xx}=\eta^+\partial_xv_x-\eta^-\partial_zv_z-P_h,\nonumber\\
&&\Sigma_{zz}=\eta^+\partial_zv_z-\eta^-\partial_xv_x-P_h,\nonumber\\
&&\Sigma_{xz}=\eta(\partial_zv_x+\partial_xv_z).
\end{eqnarray}

Neglecting the fluctuations, we see that the steady-state solution reads as:
\begin{equation}
 v_{z}^\text{s}(z)=v_\text{g}e^{z/\lambda},~~~v_\text{g}=\frac{-\Delta P}{\sqrt{\Gamma\eta^+}},
 \label{v.s1}
 \end{equation} 
 where, $\lambda= \sqrt{\eta^+/\Gamma}$ is the hydrodynamic screening length and 
  the growth velocity or the speed by which the boundary proceed 
 is denoted by $v_\text{g}$. 
 In a small region with thickness $\lambda$, just below the growing front, net flow can be observed. Beyond this layer, pressure has its equilibrium value denoted by $P_h$ and no net growth can be observed. 
 For $\Delta P > 0$ apoptosis dominates and $v_\text{g}<0$, showing that the boundary 
 moves downward.  For a system in which the growth (cell division) is dominated,  $\Delta P  < 0$ and this corresponds to 
 positive growth velocity $v_\text{g}>0$ where the boundary moves upward.  
 
 To see how the growing front remains smooth,  
 we consider the fluctuations up to the first order of the height function $b(x,t)$.
 We consider a traveling wave pattern as $b(x,t)={\tilde b}e^{i(q_x x-\omega t)}$ and investigate the response of the system. 
 Furthermore, we consider the following ansatz for the velocity profile:
 $$\delta {\bf v}(x,z,t)=\delta{\tilde {\bf v}}e^{i(q_xx-\omega t)+kz},$$
 where $k^{-1}$ shows the depth within which the fluctuations penetrate into the system.
 Inserting the above velocity pattern into the equations, we will arrive at the following equations: 
 \begin{equation}
 \begin{bmatrix}
 (\eta k^2- \eta^+ q_x^2-\Gamma ) & i\zeta kq_x\\
 i\zeta kq_x  & (\eta^+ k^2  -\eta q_x^2-\Gamma )
 \end{bmatrix}
 \begin{bmatrix}
 \delta{\tilde v}_x \\
 \delta{\tilde v}_z
 \end{bmatrix}=0.
 \end{equation}
 The non-trivial solution to the above set of equations results in two possible values for $k$. 
 At the limit of very small wave numbers for the fluctuations ($q\rightarrow 0$), we expect to have $k=\lambda^{-1}$. 
 As a result of this requirement, the solution with $k=\sqrt{q^2+\lambda^{-2}}$ is acceptable.  
Now, we investigate the boundary conditions to examine the spectrum of fluctuations. Putting the above solutions into the boundary conditions,  the dispersion relation will read as:
\begin{equation} 
-i\omega = -\frac{{\gamma_s}\lambda q_x^2}{\eta^+}
\left( 1-\frac{v_\text{g}}{2\gamma_{s}}(3\eta-\zeta) \right),
\label{dispersionscalar}
\end{equation}
As we expected, all fluctuations relax to zero for a homeostatic state with $v_\text{g}=0$.  
Far from the homeostatic state where the system grows, the flat boundary can be unstable depending on the parameters.   For 
$3\eta>\zeta$, we see that a  flat growing  boundary ($v_\text{g}>0$) is unstable for 
$v_\text{g}\ge 2\gamma_s/(3\eta-\zeta)$. On the other hand, the flat growing boundary is always stable for $\eta<\zeta/3$. 
In Fig.~\ref{fig2}, in terms of 
$\bar{\zeta} = \zeta/\eta^+$ and ${\bar v}_\text{g}=\eta^+v_\text{g}/\gamma_s$, we have investigated the possible behavior of the 
growing boundary. 
{This instability criterion can be understood as a competition between surface tension and growth. 
The activity corresponding to growth amplifies the surface undulations whereas, the surface tension provides a restoring force. 
The underlying mechanism for the instability takes its roots in the fact that the variation in the growth rate in our homeostatic model, 
would be roughly proportional to the shape variations given by function $b(x,t)$. 
As a result of shape fluctuations in the moving front, a local protrusion on the growing front would experience a higher growth rate 
which will eventually result in shape instability.

\begin{figure}[h!] 
 \begin{center}
 \includegraphics[width=0.40\textwidth]{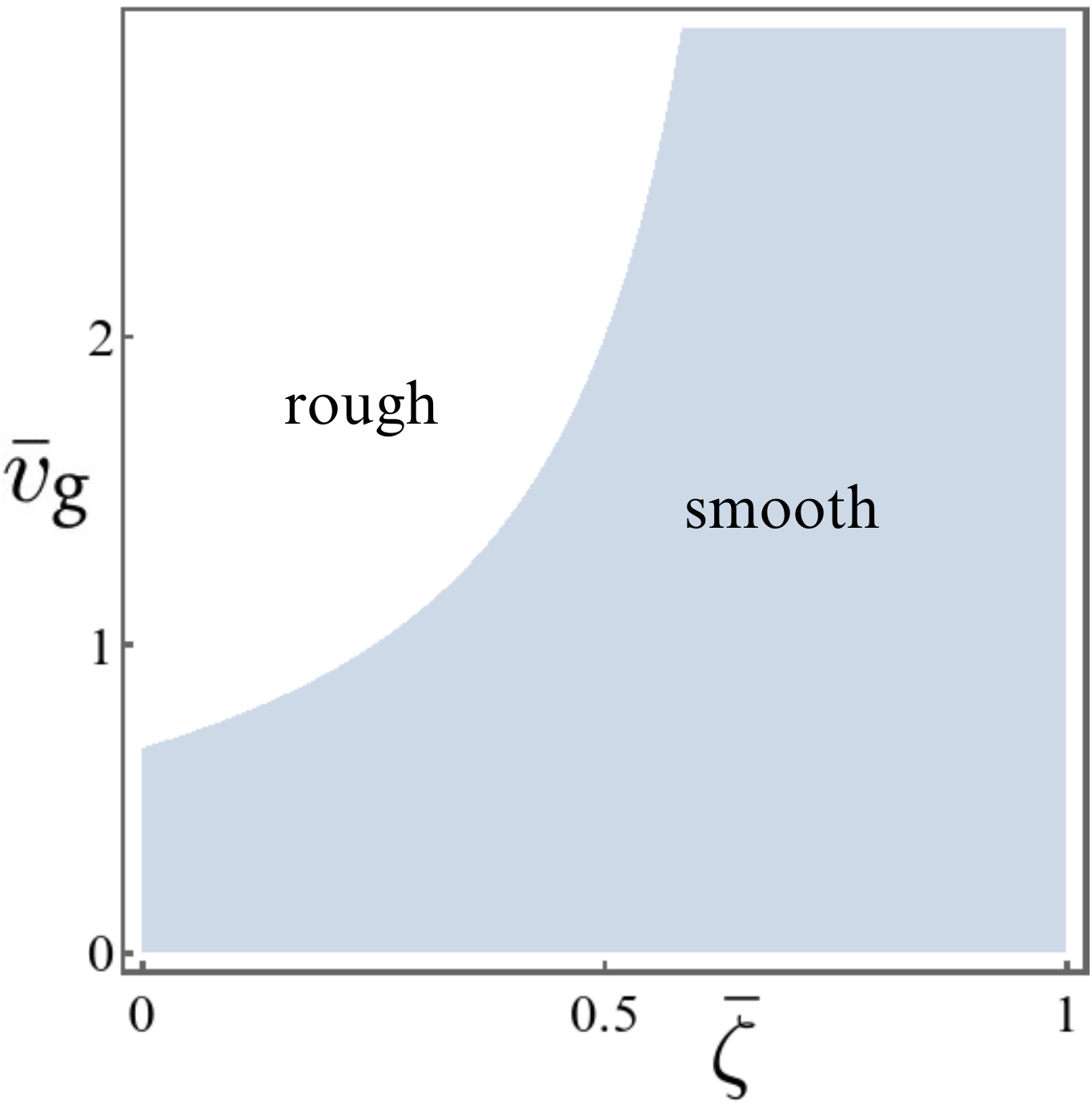}
 \caption{Phase diagram of a growing boundary in a scalar system is plotted in terms of growth activity ${\bar v}_\text{g}$ 
 and $\bar{\zeta}=\zeta/\eta^+$. The smooth phase corresponds to the case where the flat boundary is stable and 
 the rough phase corresponds to the case where the rough boundary is the stable solution.} 
\label{fig2}  
  \end{center}
\end{figure}

\section{Vectorial system (bulk)}
From now on we move to the vectorial system where orientational order plays an important role.  
In this case 
the activity parameter $a$, growth parameter $v_\text{g}$ and 
 $\beta$, the parameter reflecting the growth-mediated torque, contribute to the dynamics. 
To study the dynamics in the bulk, consider an infinite system with coarse-grained fields given by  ${\bf n} = {\bf n}_{0}$, $P^\text{g} = P_\text{h}$ and 
${\bf v} = 0$. Denoting the fluctuations by $\delta{\bf n}$ and $\delta{\bf v}$,  we linearize the dynamical equations and neglect the effect of inertia to reach the following equations for perturbative fields:
\begin{eqnarray}
		- \Gamma \mathbf{\delta v} &&- \zeta (\mathbf{q}\cdotp \mathbf{\delta v})\mathbf{q}- \eta q^{2}\mathbf{\delta v}+
		\dfrac{i}{2}(1-A)(\mathbf{q}\cdotp \delta \mathbf{h})\mathbf{n}_{0}\nonumber\\
		&&-\dfrac{i}{2}(1+A)(\mathbf{q}.\mathbf{n}_{0})\delta\mathbf{h} -i a (\mathbf{q}\cdotp \mathbf{\delta n})\mathbf{n}_{0}- ia (\mathbf{q}\cdotp \mathbf{n}_{0})\mathbf{\delta n}=0,
		\nonumber\\
-i\omega&&\delta{\bf n}= \frac{i}{2}(1+A)(\mathbf{q}\cdotp \mathbf{n}_{0})\mathbf{\delta v}-\frac{i}{2}(1-A)(\mathbf{\delta v}\cdotp \mathbf{n}_{0})\mathbf{q}+\frac{1}{\gamma}\delta\mathbf{h}\nonumber\\ 
		&&-iA(\delta{\bf v}\cdot{\bf n}_0)({\bf q}\cdot{\bf n}_0){\bf n}_0+\frac{\beta  \zeta}{\gamma} A^2(\mathbf{q}\cdotp \mathbf{\delta v})(\mathbf{q}-(\mathbf{q}\cdotp\mathbf{n}_{0})\mathbf{n}_{0}),
		\label{bulknematic}\nonumber
\end{eqnarray}
where wave vector and frequency of the perturbations are denoted by ${\bf q}$ and $\omega$, respectively. 
We denote by $\theta_0$, the angle between wave vector and director field ${\bf n}_0$ (see Fig.~\ref{fig3}). 
{Furthermore  $\delta{\bf h}/K = -({\bf q}-\mathbf{n}_0(\mathbf{n}_0\cdotp {\bf q}))({\bf q}\cdotp\mathbf{\delta n})-
(\mathbf{n}_0\cdotp {\bf q})^{2}\mathbf{\delta n}$.} 
The solution to the above equation in an unbounded space gives the dispersion relation as:
\begin{equation}
-i\omega=\tau_\text{r}^{-1}-iv({\bf q},{\bf n}_0)q,
\end{equation}
where relaxation time and group velocity of the perturbations are given by:
\begin{eqnarray} 
&&\tau_\text{r}^{-1}=\frac{a}{{\Gamma}/q^2+\eta}\left(\frac{1}{2}\cos 2\theta_0+\frac{A}{4}g(\theta_0)\right)\nonumber\\
&&~~~~~-{K}_\text{e}\left({\Gamma}/q^2+\eta+\gamma/4+\frac{A}{2}\gamma\cos2\theta_0+A^2\gamma g(\theta_0)\right)\nonumber\\
&&v({\bf q},{\bf n}_0)= \frac{ \beta  \zeta A^2  \sin 2\theta_0}{\gamma(\eta^+ +\Gamma/q^2)}\left(a-2AKq^2\right),
\end{eqnarray}
here 
${K}_{\text{e}}=\frac{K q^2}{\gamma(\eta+{\Gamma}/q^2)}$, 
and function $g(\theta_0)$ is given by:
$$
g(\theta_0)=\frac{2{\Gamma}/q^2+2\eta+\zeta(1+\cos(4\theta_0))}{\eta^++{\Gamma}/q^2}.
$$
Reflected from the first term in $\tau_\text{r}$, for  $a>0$ ($a<0$), splay (bend) fluctuations tend to initiate hydrodynamic instability for geometrically symmetric swimmers ($A=0$) \cite{ramaswamy1}. Bend and splay elastic energy can stabilize both modes of perturbations and this is shown in the 
second term in $\tau_\text{r}$ \cite{ramaswamy1}. 
The interesting physics is in the real part of frequency $\omega_\text{r}=v({\bf q},{\bf n}_0)q$. When the hydrodynamic instabilities are stabilized by nematic elastic energy or other stabilization mechanisms \cite{mehrana}, stable traveling waves can propagate in the system with corresponding group velocity given by $v$. Propagation of such active waves is directly related to the parameter $\beta$, the
growth-mediated torque. Active waves can be observed in systems that either contain motile particles ($a\neq 0$, $A\neq 0$) or contain non-motile particles with finite rotational elasticity ($K\neq 0$, $A\neq 0$). For a system with nematic order, the velocity of such propagating modes crucially depends on the direction of propagation. The maximum velocity of propagating active waves corresponds to the case where the wave vector has an angle $\theta_0=\pi/4$ with the nematic direction. 
Fig.~\ref{fig3}, shows a snapshot of the active wave which is propagating in a direction with its maximum velocity.  
It is interesting to note that  
pure bend and splay waves ($\theta_0=0$ and $\theta_0=\pi/2$) can not propagate. In addition to the director wave, one can consider 
this traveling wave as a pressure or growth wave.  Local fluctuations in growth rate can propagate in the system. More 
interestingly is the direction of propagation which is a right-moving wave in the sense that fixing an angle $\theta_0$, the waves can only propagate in $+{\hat q}$ direction. In Fig.~\ref{fig3}, we have presented an intuitional picture that can reveal the 
physics behind this active wave. As seen in this picture, in a locally ordered nematic phase, a small fluctuation in the direction of a 
particle can produce a hydrodynamic flow. Divergence of this excess flow initiates a pressure gradient and subsequently 
gives rise to a gradient in the 
growth rate. Then,  growth-mediated torque will eventually promote the fluctuations to propagate. 
\begin{figure}
 \begin{center}
 \includegraphics[width=0.40\textwidth]{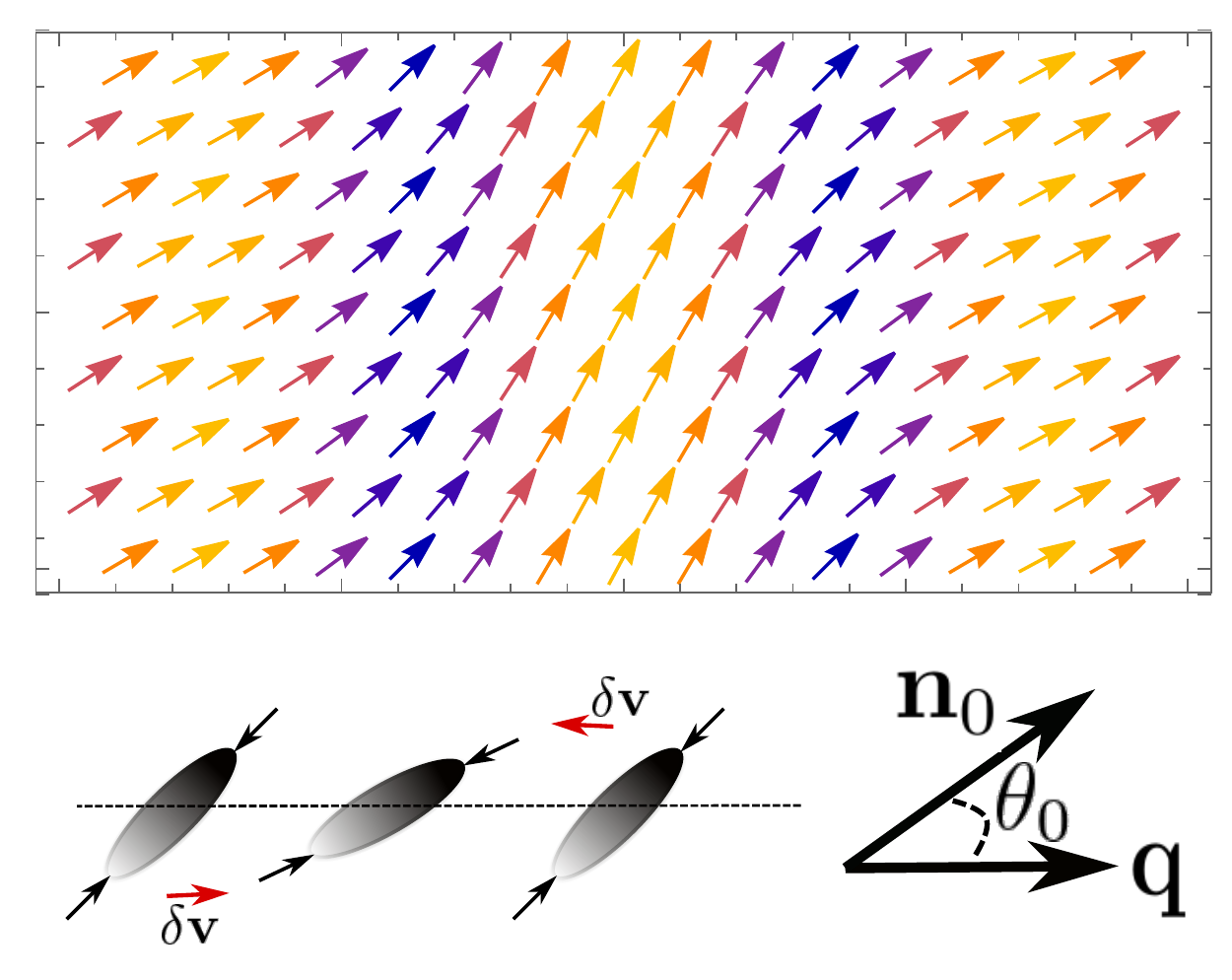}
      \caption{ up: In terms of orientation and local growth rate (encoded in the color of arrows), the active traveling wave is shown. We have chosen an angle 
      $\theta_0=\pi/4$ corresponds to a wave with maximum velocity.  Here the local growth density is encoded in the 
      color by which the orientation vectors are drawn.  down: To see how a perturbation in the director field can propagate, 
      we have applied a small orientational fluctuation to the middle cell.  Cells are assumed to be contractile (puller) 
      and their corresponding 
      flow pattern is shown by black arrows. As a result of a small orientational fluctuation, a local velocity denoted by 
      $\delta{\bf v}$ will emerge. In the homeostatic picture,  ${\bf q}\cdot\delta{\bf v}$ will give a pressure difference 
      and it eventually gives a gradient in growth rate.  Taking into account the growth-mediated torque, this will eventually 
      provide an active 
      source for traveling waves. 
       }
\label{fig3}  
  \end{center}
\end{figure} 

\section{Vectorial system (boundary)}
Having studied the bulk properties of an active nematic system, we now consider a system that is bounded by a rigid wall at $z= -\infty$ and a freely growing boundary at $z=0$, see Fig.~\ref{fig1}.  
At very long times the system reaches a steady state in which the boundary moves with velocity $v_\text{g}$. 
The steady-state velocity profile is similar to the scalar case given in Eq.~\ref{v.s1}, 
with the growth velocity of the  boundary that is replaced by:
\begin{equation}
{v}_\text{g}=\frac{-\Delta P+a}{\sqrt{\Gamma\eta^+}}.
\end{equation}
This steady state corresponds to the case where all elongated particles are perpendicular to 
the moving front. 
To consider the dynamics of fluctuations we put ${\bf n}={\hat z}+\delta\theta(x,z,t){\hat x}$ and denote the  
shape of the interface by the function $b(x,t)={\tilde b}e^{i(q_xx-\omega t)}$. Similar to the scalar case, we assume that all the bulk fields behave like:
$\delta {\bf v}(x,z,t)=\delta{\tilde {\bf v}}e^{i(q_xx-\omega t)+kz}$. Putting this information in the equations, we arrive at the following equation for the amplitudes:
\begin{equation}
\begin{bmatrix}
 \eta
k^{2}-\eta^+ q_x^{2}-\Gamma & i q_x \zeta k & 
-k(a+d^+)\\
i q_x\zeta k & 
\eta^+k^{2}-\eta q_x^{2}-\Gamma& -i
 q_x (a+d^-) \\
 (\dfrac{k}{2}+\beta_1 q_x^{2}) & -i q_x(\dfrac{1}{2}+\beta_1k) & 
(\beta_1 \dfrac{v_\text{g}}{\lambda^{2}}+\frac{d}{\gamma})
\end{bmatrix}
\begin{bmatrix}
\delta {\tilde v}_x\\
\delta {\tilde v}_z\\
\delta {\tilde \theta}
\end{bmatrix}=0,\nonumber
\end{equation}
where, $d^\pm=\frac{A\pm1}{2}d$ with $d=K(k^2-q^2)$, $\beta_1=\beta\zeta \gamma^{-1}A^2$. The solution to the above homogeneous system of equations, reveals the penetration depth $k^{-1}$. Note that we assumed that the rotational dynamics of the director field are fast so we neglected the unsteadiness of the director field in the bulk.   
Putting the solutions in the boundary conditions, Eqs.~\ref{BC1} and \ref{BC2}, will give us:
\begin{eqnarray}
&&\delta {\tilde \theta}=-iq_x{\tilde b},~~~~\delta {\tilde  v}_z+{\tilde b}\frac{v_\text{g}}{\lambda}=-i\omega {\tilde b},\nonumber\\
&&-iq_x\eta^-  \delta\tilde{v}_{x}+k\eta^+ \delta\tilde{v}_{z}+
\left(\gamma_s q_x^{2}+\eta^+\frac{v_\text{g}}{\lambda^2}\right){\tilde b}=0,\nonumber\\
&& \eta k\delta\tilde{v}_{x}+i\eta q_x\delta\tilde{v}_{z}+2iq_x\frac{v_\text{g}}{\lambda}{\tilde b}-d^-\delta{\tilde \theta}=0.
\label{bc4}
\end{eqnarray}
Again, the above equations can be considered as a set of homogenous equations incorporating the field 
amplitudes as unknown variables. Looking for non-zero solutions for variables, we will obtain a relation that reveals the frequency of oscillations.  Up to the leading order of wave vector $q$, the dispersion relation reads as:
\begin{equation}
-i\omega= \frac{\gamma_s\lambda q_x^2}{\eta^+}\left({\bar v}_g(\xi+\frac{\eta^-}{\eta^+})+\frac{l_K}{2}(A-1)(1-\frac{\zeta}{\eta})-1\right),
\end{equation}
where,
\begin{eqnarray}
&&\xi=[2{\bar \gamma}+l_Kl_a((1+A){\bar \gamma}-4({\bar \eta}^2+{\bar \zeta}^2+{\bar \eta}))\nonumber\\
&&~~~~~~~~~~+2
{\bar \zeta}A^3l_K l_a l_\beta(1-\frac{\eta^-}{\eta^+})\nonumber\\
&&~~~~~~~~~~+4{\bar \zeta}A^2 l_\beta({\bar \zeta}-{\bar \zeta}^2{\bar v}_gl_a+{\bar \zeta}(1-{\bar \eta}{\bar v}_gl_a))
]\nonumber\\
&&~~~\times[4{\bar \gamma}-8A^2 l_al_\beta{\bar v}_g{\bar \zeta}^2+2l_Kl_a(-4{\bar \zeta}+(1+A){\bar \gamma})]^{-1},\nonumber
\end{eqnarray}
where dimensionless variables are defined as: $l_K=(\ell_K/\lambda)$, $l_a=(\ell_a/\lambda)$, 
 $l_\beta=\ell_\beta/\lambda$ with $\ell_K=(K/\gamma_s)$, $\ell_a=(\gamma_s/a)$, 
 $\ell_\beta=\beta$  and ${\bar \gamma}=\gamma/\eta^+$. 
 
 Before analyzing the growth-mediated instabilities, we note that at the limit of ${\bar v}_{g}=0$, a passive instability can be observed. As it is seen from the above relation (setting ${\bar v}_{g}=0$),  the instability can arise from a competition between surface tension and bulk elasticity. For $(A-1)(\eta-\zeta)>(2\eta/l_K)$, this will give instability. It should be noted that this instability is not a general feature of passive systems. Here, the compressibility of the fluid combined with the special choice of the ordered 
 state in which ${\bf n}_0$ is perpendicular to the boundary, triggers the instability. Local terms proportional to $n^2$ and $n^4$ in the free energy, the terms that are not considered in our model, can stabilize this passive instability.  
 
To analyze the growth-associated instabilities, we consider the case where  ${\bar v}_{g}\neq 0$. It is seen that 
the elasticity, motility, and growth-mediated torque,  contribute to the instability through their corresponding length scales denoted by $\ell_K$, $\ell_a$, and $\ell_\beta$, respectively. 
Among these different parameters, we investigated the phase diagram of the system in terms of the speed of 
growth ${\bar v}_g$, particle asymmetry $A$ and the strength of growth-mediated torque $\beta$.  
 The phase diagram of the system for a special choice of parameters is plotted in fig.~\ref{fig4}. It is shown how the asymmetry parameter of particles $A$, competes with growth parameter ${\bar v}_g$ to result in the either smooth or rough 
 boundary for the system. Similar to the scalar case, there is always a threshold growth speed ${\bar v}_g$, beyond which the moving front gets roughness. Slow growth with speed less than this threshold speed will result in a flat and smooth interface. 
Fig.~\ref{fig4} shows how this threshold velocity behaves as a function of particle asymmetry $A$ and growth-mediated parameter $\beta$. 
\begin{figure}[h!] 
 \begin{center}
 \includegraphics[width=0.45\textwidth]{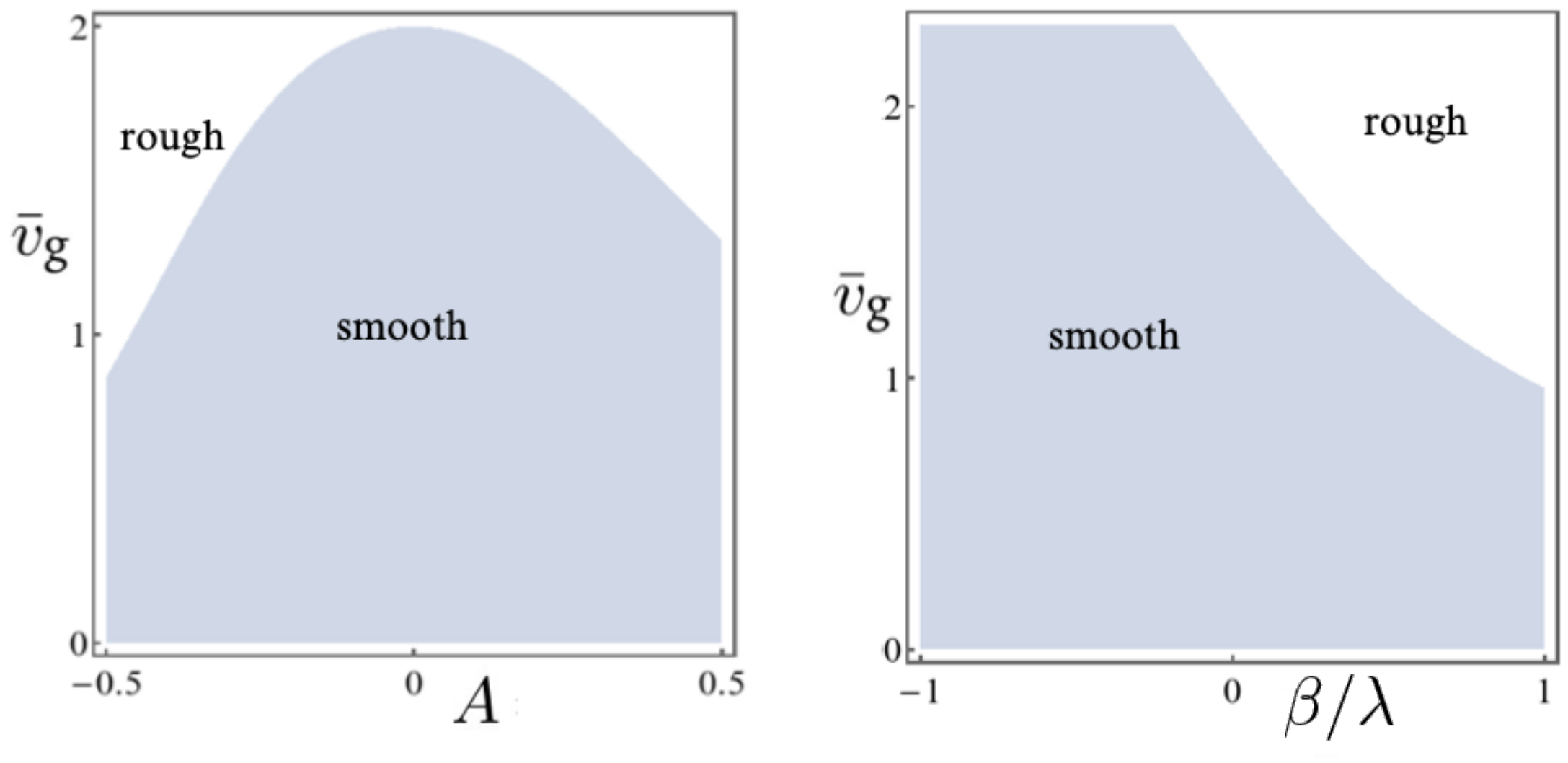}
 \caption{
Phase-diagram of a growing boundary in a vectorial system is plotted in terms of growth activity ${\bar v}_\text{g}$ 
 and asymmetry parameter $A$. 
 The smooth phase corresponds to the case where the flat boundary is stable and 
 the rough phase corresponds to the case where the flat boundary is not the stable solution. Here 
.(left:  $l_K=l_a=l_\beta={\bar \gamma}=1,\bar{\zeta}=0.5$; right:$l_K=l_a=\bar{\gamma}=1,\bar{\zeta}=0.5,A=0.7$)}
\label{fig4}  
  \end{center}
\end{figure}

\section{Discussion}
We studied the dynamical instabilities in a growing system. Our model takes into account four different length scales. 
As a result of friction with a substrate, the hydrodynamic interactions are screened and $\lambda$ shows the corresponding screening length. The elasticity of the bulk introduces another length 
scale that is denoted by $\ell_K$. Two other length scales correspond to the motility and growth-mediated torque, those are denoted by $\ell_a$ and $\ell_\beta$, respectively. 
Our analysis shows that for a scalar system, in the case where the orientational degrees of freedom 
is neglected, all dynamical behavior is limited to a boundary layer with thickness $\lambda$ near the free interface 
of the system. All variations at the bulk will rapidly decay, but the fluctuations near boundaries can result in shape 
instabilities in the interfaces. In contrast, for the vectorial case, the case where the rotational degrees of particles play an important role, nontrivial results can be observed both at the bulk and interface. As a result of a phenomenological growth-mediated torque, we observed an active wave that can propagate in the bulk. The speed of this active wave is roughly proportional to $(\beta\zeta a/\gamma \Gamma)q^2$.  Increasing either the friction with the substrate or the rotational friction of particles will result in a decrease in the propagation speed. 
This wave can be considered as a wave pattern on the pressure field in the system. As the fluctuation in the pressure is proportional to the growth rate, the active wave can also be thought of as a propagating wave in the pattern of growth rate. We are not aware of any real observation of such a wave but we think this might influence the overall dynamics of growing colonies. In addition to the bulk properties, we also studied the interface instabilities in the vectorial case.

\section{Acknowledgement} Useful discussions with R. Golestanian and F. Julicher and helps received from M. Setoudeh at the early stage of the work are acknowledged. 

\nocite{apsrev41Control}
\bibliography{Ref.bib,revtex-custom}

\end{document}